\documentclass[12pt,epsf]{article}
\usepackage{epsfig}

\newcommand{\onefigure}[2]{\begin{figure}[htbp]
\begin{center}\leavevmode\epsfbox{#1.eps}\end{center}\caption{#2\label{#1}}
\end{figure}}
\newcommand{\setfigure}[2]{\begin{figure}[htbp]
\begin{center}\leavevmode\epsfxsize=5in\epsfbox{#1.eps}\end{center}\caption{#2\label{#1}}
\end{figure}}
\newcommand{\twofigures}[3]{\begin{figure}[htdp]
\centering \leavevmode\epsfxsize=2.5in\epsfbox{#1.eps}
\leavevmode\epsfxsize=2.5in\epsfbox{#2.eps} 
\caption{{
#3}\label{#1}}
\end{figure}}

\renewcommand{\thanks}[1]{\footnote{#1}} 

\newcommand{\be}{\begin{equation}}
\newcommand{\ee}{\end{equation}}
\newcommand{\bea}{\begin{eqnarray}}
\newcommand{\eea}{\end{eqnarray}}

\begin{document}

\pagestyle{empty}

\bigskip\bigskip
\begin{center}
{\bf \large Construction of a Penrose Diagram for a Spatially Coherent Evaporating Black Hole}
\end{center}

\begin{center}
Beth A. Brown\footnote{e-mail address, beth.a.brown@nasa.gov}\\
James Lindesay\footnote{e-mail address, jlslac@slac.stanford.edu} \\
Computational Physics Laboratory \\
Howard University,
Washington, D.C. 20059 
\end{center}
\bigskip

\begin{center}
{\bf Abstract}
\end{center}
A Penrose diagram is constructed for an example
black hole that evaporates at a steady rate as measured
by a distant observer, until the mass vanishes, yielding
a final state Minkowski space-time.  Coordinate
dependencies of significant features, such as the
horizon and coordinate anomalies, are clearly
demonstrated on the diagram.  The large-scale
causal structure of the space-time is briefly
discussed.
\bigskip \bigskip \bigskip

\setcounter{equation}{0}
\section{Introduction}
\indent

Often, one's intuitive feel for the behavior of a dynamic black hole is
guided by the assumption that
the quasi-static dynamics are dominated by small
modifications on the
geometry of the related static black hole.  However, there is some
evidence\cite{JLMay07} that the asymptotic behaviors of dynamic
geometries differ from those of static geometries. 
Indeed, one expects that the initiation and final
evaporation processes of black hole dynamics
should involve spatial coherence on scales comparable
to those describing the medium scale structure of the
space-time.  It
is therefore of interest to examine examples of temporally
dynamic geometries, along with their correspondence
to static space-times.

The non-orthogonal temporal coordinate associated with the
river model of black holes\cite{rivermodel} and related
descriptions of dynamic horizons\cite{Nielsen} has been
shown to provide a convenient parameter for describing the
evolution of a black hole without introducing new
physical singularities due to coordinate anomalies away from
the point of symmetry $r=0$.  Such a time coordinate
mixes sufficiently with the radial spatial coordinate in a
manner that gives some spatial coherence to the
``nearby" dynamics described using this temporal parameter.
One expects that any coherent processes associated with the
radiations of evaporating black holes will have
spatial scales comparable to the radial mass scale
(e.g. Schwarzschild radius) of the geometry.  The initiation, growth
by accretion, evaporation, and finality of a dynamic black hole
should have aspects of spatial coherence on the scales
associated with those local ``microscopic" dynamics.

This paper develops the Penrose diagram of a geometry with a
steady rate of evaporation as introduced in a previous paper\cite{JLMay07}
in order to examine the large-scale causal structure of an
example dynamic black hole.  The particular example is chosen
because classical and quantum
solutions can be obtained in a straightforward manner.

\setcounter{equation}{0}
\section{Geometry of an Evaporating Black Hole
\label{nonsingularhorizon}}

\subsection{Form of the metric}
\indent

For the present examination, the
space-time metric will take the form
\be
ds^2 = -\left (1-{R_M (ct_a) \over r} \right ) (dct_a) ^2 +
2 \sqrt{{R_M (ct_a) \over r}} dct_a \, \, dr + dr^2 + r^2 \, d \omega ^2
\label{metric}
\ee
where $d\omega^2 \equiv d\theta^2 + sin^2 \theta \, d\phi^2$. 
This space-time asymptotically corresponds with
Minkowski space $(ct_a,r,\theta,\phi)$ in a manner
similar to (but not necessarily identical to)
the behavior of a Schwarzschild geometry.  Therefore, $ct_a$
represents the time coordinate of an asymptotic observer. 
As demonstrated in
reference \cite{JLMay07}
the Ricci scalar
\be
\mathcal{R}={3 \dot{R}_M \over 2 r^2} \sqrt{r \over R_M}
\ee
is non-singular away from a physical singularity
at the origin, and vanishes for a static radial mass
scale $\dot{R}_M\rightarrow 0$. 
This means that the invariant curvature (and any
related invariant physical parameter) is
nowhere singular away from the origin.  Any
coordinate singularities manifest only in components
unique to that particular coordinate representation, and do
not represent singularities in the physical content of the
space-time.

\subsection{Evolution of the horizon
\label{horizonsection}}
\indent

The  radial null surfaces for the metric specified in Eq. \ref{metric}
satisfy
\be
{d r_\gamma \over dct_a} = - \sqrt{R_M \over r_\gamma} \pm 1.
\ee
Outgoing photons (where $r_\gamma$ increases with $ct_a$)
traverse trajectories that correspond to the upper sign.
The radial coordinate corresponding to the horizon is given by
the particular null surface proportional to the radial mass scale
$R_H={R_M \over \zeta_H}$, where
\be
{d R_H \over dct_a} = - \sqrt{R_M \over R_H} + 1 \quad , \quad 
\dot{R}_M = 
\zeta_H \, \left(1-\sqrt{\zeta_H} \right ) \quad ,  \quad
R_H = {R_M \over \left (1 - \dot{R}_H  \right ) ^2} \quad .
\label{horizon}
\ee
For the case presently being examined, the horizon is seen to be always
within the radial mass scale $R_H < R_M$.

\subsection{Diagonalization of the metric form
\label{diagsection}}
\indent

The construction of orthogonal temporal-radial coordinates
presents a more direct intuitive representation for the characterization
of physical dynamics.  If one attempts a coordinate transformation
of the form
\be
d ct_a = A(ct_a,r) \, d ct_D + \Delta(ct_a,r) \, dr \quad , \quad
dr = dr_D ,
\label{timetransform}
\ee
the function $\Delta$ can be chosen to algebraically diagonalize
the metric in Eq. \ref{metric} if it is of the form
\be
\Delta(ct_a, r) = { \sqrt{ {R_M (ct_a) \over r} } \over 1 - { R_M (ct_a) \over r}  }.
\label{Delta}
\ee
Observers using coordinates $(ct_D,r,\theta,\phi)$,
including the diagonal temporal coordinate
$ct_D$, describe no frame-dragging affects
in this geometry. 
The diagonalized metric then takes the form
\be  
ds^2 = -(1- {R_M(ct_a) \over r} ) A^2 (ct_a,r) \, (dct_D) ^2 + 
{dr^2 \over 1 - {R_M(ct_a) \over r} } +
r^2 (d \theta ^2 + sin^2 \theta \, d\phi ^2) ,
\label{diagonalmetric}
\ee
where the integrability condition on the coordinate
$ct_D$ constrains the temporal scale factor $A$, requiring
that it satisfies
\be
{\partial \over \partial r} {1 \over A}  =
-{\partial \over \partial ct_a} { \Delta \over A} .
\label{integrability}
\ee
If $\ddot{R}_M=0$, a solution can be demonstrated for the coordinate
transformation.  The reduced coordinate $\zeta$ will
be defined by
\be
\zeta(ct_a,r) \equiv {R_M (ct_a) \over r}. 
\ee
The temporal scale factor $A$ is assumed to approach unity for vanishing $\dot{R}_M$,
giving the usual static Schwarzschild coordinates.  
The coefficient $A(\zeta)$ then satisfies
\be
\zeta {\partial \over \partial \zeta} {1 \over A(\zeta)} =
\dot{R}_M {\partial \over \partial \zeta} {\Delta (\zeta) \over A(\zeta)} \quad , \quad
A(\zeta) =  \,exp  \int_{\zeta_o} ^{\zeta} \left [
{{\partial \Delta(\tilde{\zeta}) \over \partial \tilde{\zeta}} \, \dot{R}_M \, d \tilde{\zeta} \over
\Delta(\tilde{\zeta}) \, \dot{R}_M - \tilde{\zeta} }\right ]  ,
\label{Aofzeta}
\ee
where $\zeta_o$ is the reduced coordinate of correspondence with the
Schwarzschild metric.
The behavior of this function is shown in Fig. \ref{ScaleFac}. 
\onefigure{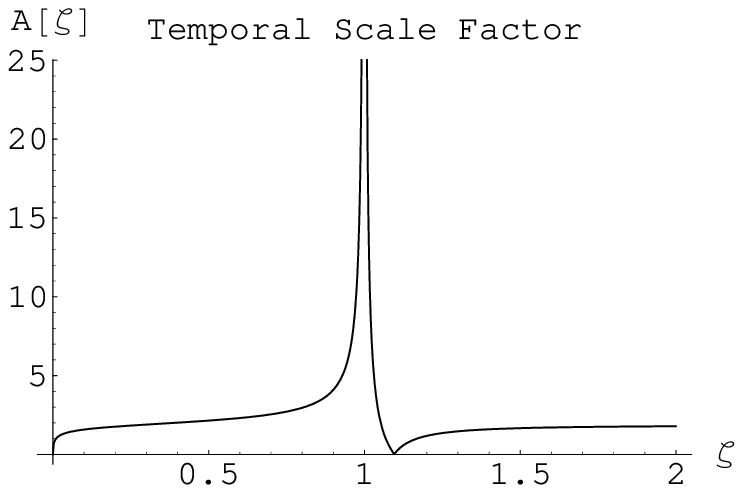}{Scale Factor for an evaporating black hole with $\dot{R}_M=0.1$}
The temporal scale factor is seen to behave in a manner of interest for
several values of $\zeta$ in the figure. 
One of the zeroes of the temporal scale factor
$A(\zeta_A)=0$ corresponds to an anomaly in the diagonal time
coordinate $ct_D$, while the singular behavior in $A$ near $\zeta \sim 1$ corresponds to the coordinate
anomaly in $ct_a$ associated with the radial mass scale $R_M$.  The zero in the
temporal scale factor for $\zeta=0$ dis-allows the use of $r \rightarrow \infty $
for correspondence of $ct_D$ with the temporal coordinate
of asymptotic Schwarzschild geometry.

The form of the diagonalized time can be integrated from the analyticity
requirement associated with the coordinate transformation.  Suppose that
the form of this coordinate is given by
\be
ct_D (ct_a,r)=R_M(ct_a) \, F(\zeta) .
\label{diagonaltime}
\ee
Substitution into Eq. \ref{timetransform} gives conditions
\be
\zeta^2 F'(\zeta)={\Delta (\zeta) \over A(\zeta)} \quad , \quad \dot{R}_M \left [
F(\zeta) + \zeta F'(\zeta) \right ] = {1 \over A(\zeta)}
\ee
which can immediately be solved for the function $F(\zeta)$ to give
\be
ct_D (ct_a,r)=\left (  {\dot{R}_M \sqrt{\zeta} - \zeta (1-\zeta) \over
\zeta (1-\zeta) A(\zeta)} 
\right ) {R_M(ct_a) \over \dot{R}_M} .
\label{diagonalanswer}
\ee
This diagonal form will be useful in the development of the
conformal coordinates needed to construct the Penrose
diagram for this space-time.

\subsection{Construction of conformal coordinates}
\indent

Conformal coordinates are convenient for examining the
causal structure of the space-time.  The conformal coordinates $(ct_*, r_*)$
for this space-time can be obtained from the diagonal coordinates using the transformation
\bea
B \, (dct_* -D \, dr_*) \equiv A \sqrt{1-\zeta} \, dct_D  \nonumber \\ 
B \, (-D \, dct_* + dr_*) \equiv {dr \over \sqrt{1-\zeta}} \quad . \quad
\label{transformation}
\eea
The metric then takes the form
\be
ds^2 = B^2 (1-D^2) \left [ -(dct_*)^2 + dr_*^2 \right ] + r^2 \left ( d\theta^2 + sin^2 \theta \, d\phi^2 \right ) .
\ee
It is convenient to define scaling parameters $W_\pm$ using the forms
\bea
B \equiv {1 \over 2} {1 \over \sqrt{1 - \zeta}} {W_+ + W_- \over W_+ W_-}
\nonumber \\ \nonumber \\
D \equiv {W_+ - W_- \over W_+ + W_-} \: \: \: . \quad \quad \quad
\eea
The metric can then be re-written in terms of re-scaled parameters $\tilde{W}_\pm$
defined by
\be
\tilde{W}_\pm \equiv \left ( 1 \mp \sqrt{\zeta} \right ) W_\pm
\ee
in the form
\be
ds^2 = { -(dct_*)^2 + dr_*^2 \over \tilde{W}_+ \tilde{W}_-}
+ r^2 (d\theta^2 + sin^2 \theta \, d\phi^2) .
\ee

Defining conformal light cone coordinates $du_\pm \equiv dct_* \pm dr_*$,
the transformation Eq. \ref{transformation} requires that
\be
du_\pm =
\left [ \left ( 1 \pm \sqrt{\zeta} \right ) dct_a
\pm dr
\right ]
\,  \tilde{W}_\pm (\zeta) .
\ee
The functions $\tilde{W}_\pm$ then must satisfy the integrability conditions for
the conformal coordinates
\be
-\zeta {d \over d \zeta} \left [ \left ( 1 \pm \sqrt{\zeta} \right ) \tilde{W}_\pm
\right ] = \pm \dot{R}_M {d \over d \zeta} \tilde{W}_\pm \, .
\ee
This equation can be re-written in a form that more directly demonstrates
the singular points in the equation
\be
{d \over d\zeta} \left [ \left \{ \zeta \left ( 1 \pm \sqrt{\zeta} \right )
\pm \dot{R}_M \right \} \tilde{W}_\pm \right ] =
\left ( 1 \pm \sqrt{\zeta} \right ) \tilde{W}_\pm .
\ee
The roots $\zeta_S$ of the quantity in the
brackets $\left \{ \zeta_S \left ( 1 \pm \sqrt{\zeta_S} \right )
\pm \dot{R}_M \right \}=0$ represent light-like surfaces, one of which corresponds
to the horizon.  The solutions for the re-scaled factors $\tilde{W}_\pm$ are of the form
\be
\tilde{W}_\pm (\zeta) = {\pm \dot{R}_M \over \left \{ \zeta \left ( 1 \pm \sqrt{\zeta} \right )
\pm \dot{R}_M \right \} } \, exp \left [ \int_0^\zeta { \left (
1 \pm \sqrt{\zeta '}  \right ) d \zeta '  \over
\left \{ \zeta' \left ( 1 \pm \sqrt{\zeta'} \right )
\pm \dot{R}_M \right \} }
\right ] .
\ee
An analytic form for this solution can be obtained, and
an example numerical solution is demonstrated in Fig. \ref{Wplus}.
\twofigures{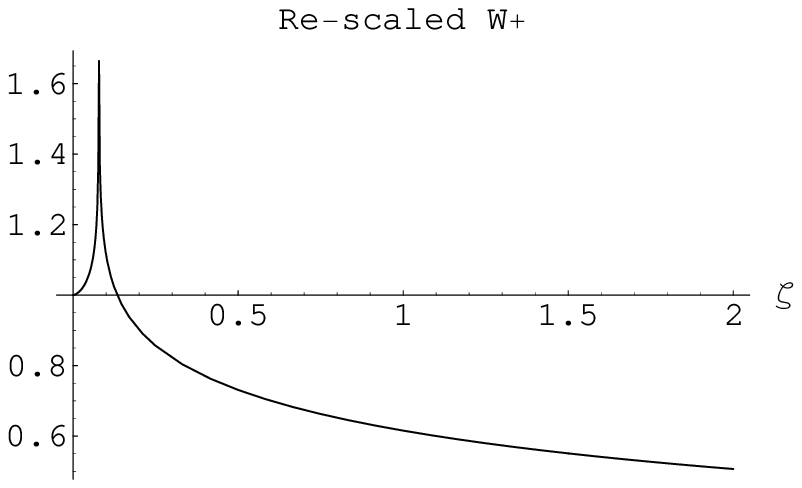}{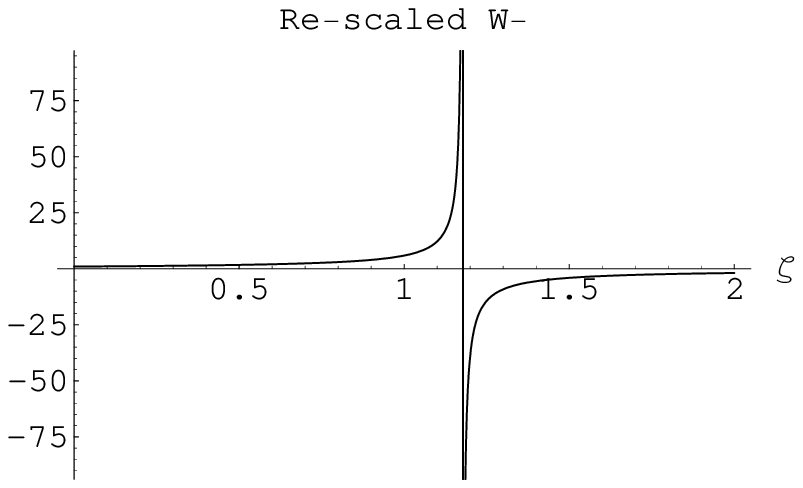}{Re-scaled transformation factors $\tilde{W}_\pm (\zeta)$
for $\dot{R}_M=0.1$}
The singular behavior in $\tilde{W}_-$ corresponds to the horizon, while that in
$\tilde{W}_+$ corresponds to an incoming light-like surface.

Utilizing the same technique used to derive Eq. \ref{diagonalanswer}, the forms of
the conformal coordinates can be integrated to obtain
\bea
ct_* = {r \over 2} \left [
\left \{ 1 + {\zeta \left ( 1 + \sqrt{\zeta} \right ) \over \dot{R}_M }  \right \} \tilde{W}_+ (\zeta) +
\left \{ -1 + {\zeta \left ( 1 - \sqrt{\zeta} \right ) \over \dot{R}_M }  \right \} \tilde{W}_- (\zeta) 
\right ] \nonumber \\ \nonumber \\
r_* = {r \over 2} \left [
\left \{ 1 + {\zeta \left ( 1 + \sqrt{\zeta} \right ) \over \dot{R}_M }  \right \} \tilde{W}_+ (\zeta) -
\left \{ -1 + {\zeta \left ( 1 - \sqrt{\zeta} \right ) \over \dot{R}_M }  \right \} \tilde{W}_- (\zeta) 
\right ] .
\label{conformal}
\eea
These equations provide the transformation between the conformal coordinates
$(ct_*,r_*)$ and the non-orthogonal coordinates $(ct_a,r)$.

\subsection{Penrose diagram of the evaporating black hole
\label{evaporation}}
\indent

In a Penrose diagram, the space-time
structure is represented using functions of
the conformal coordinates in Eq. \ref{conformal} (with
light-like surfaces always represented by lines with slope $\pm 1$)
and the entire space-time mapped onto a finite diagram.  Since
light-like surfaces are represented in a simple way, causal
relationships can be determined in such
diagrams in a straightforward manner.  Each point $(ct_a,r)$ in the Penrose
diagram developed here will represent the surface of a sphere of area $4 \pi r^2$
at distant observer time $t_a$.

\setfigure{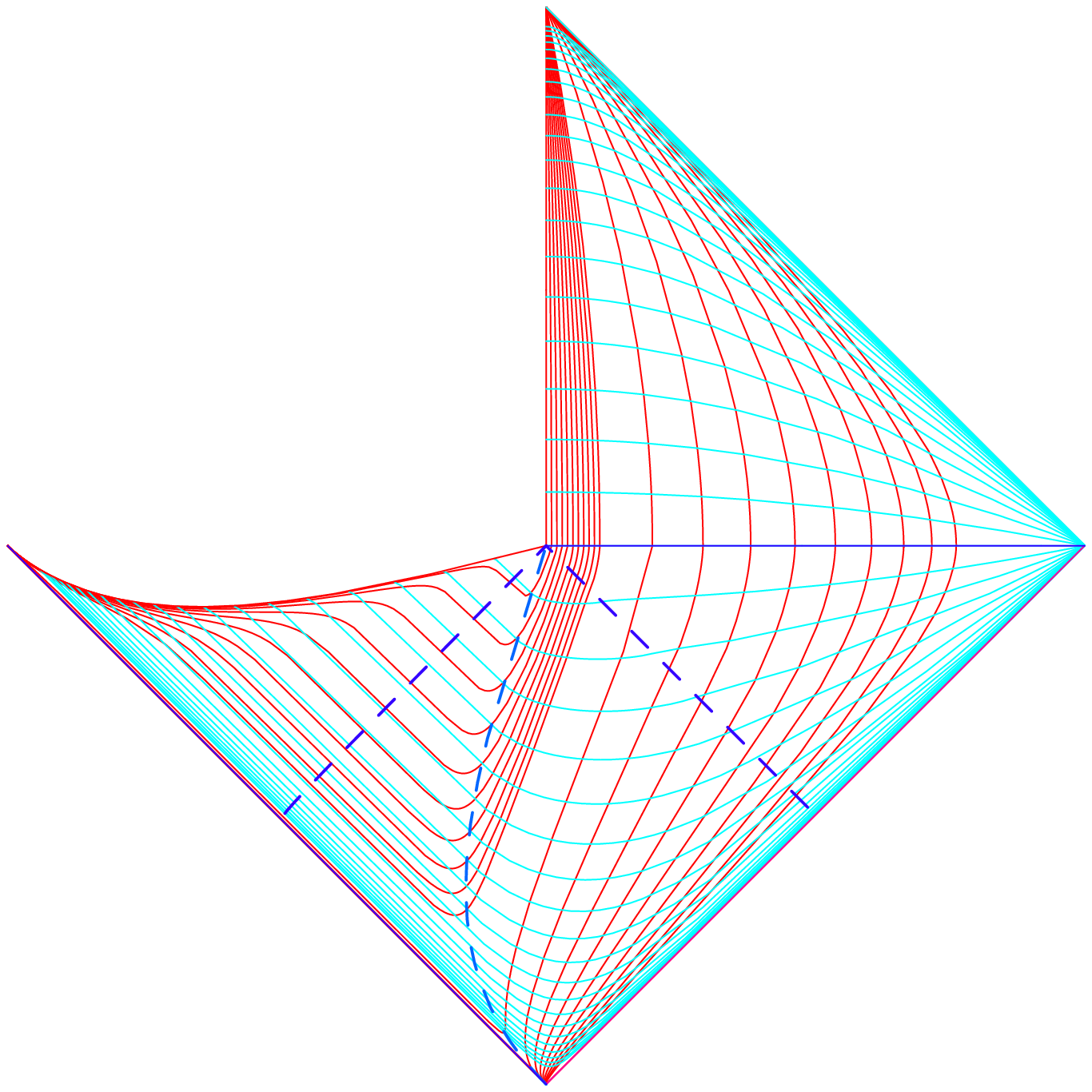}{Penrose diagram for black hole that evaporates steadily
until it reaches zero mass at $t_a=0$}

The Penrose diagram in Figure \ref{Penrose} demonstrates
the expected global structure of this spherically symmetric black hole that
evaporates at a steady rate of change in the radial mass
scale $r=R_M (ct_a)$ with respect to the distant observer's time coordinate $ct_a$.  In the distant
past, the area of the horizon of the black hole is indefinitely large, $R_H \rightarrow \infty $
as $t_a \rightarrow -\infty $.  The diagram uses hyperbolic tangents of a
scaled multiple of the conformal coordinates in Eq. \ref{conformal} for
the construction. 
In the diagram, the red curves that are time-like in the right hand
regions represent curves of constant $r$, originally graded from
$r=0$ in tenths, then in multiples of the chosen scale. 
The curves of constant radial coordinate $r$ all originate at the far left corner of the
diagram representing $t_a=-\infty$ and terminate at the uppermost corner
representing $t_a=\infty $.  The green curves that are space-like in the
right hand regions represent curves of constant $ct_a$ graded in multiples
of the given scale.  All constant $ct_a$ curves originate on the curve $r=0$
and terminate at the far right corner of the diagram representing $r=\infty $. 
The $ct_a=constant$ and $r=constant$ curves, although not
orthogonal, do serve as valid coordinates, each intersecting at only
one point on the diagram.

The diagram has several regions and features
of interest.  The right triangular region
in the upper right region of the diagram bounded by the red line $r=0$ on the left,
the solid blue line $ct_a=0$ on the bottom, and $r=\infty , ct_a=\infty $ as the hypotenuse,
corresponds to the Minkowski space-time after the completion of the
(coherent) evaporation at $t_a=0$.  For $t_a <0$ the red curve $r=0$ becomes
space-like and serves as the topmost bound of the interior region
of the black hole (the bottom left side of the diagram),
taking the curved form demonstrated between
$t_a=-\infty $ and $t_a=0$.  In the space-time region with
significant curvature, the curve $r=0$ tracks a physical singularity
with a strength related to the shrinking mass scale. 
The \emph{horizon} is a light-like surface representing
the outermost set of out-going photons that will eventually hit the
fading singularity.
On the Penrose diagram, the horizon is represented
by the dashed blue outgoing light-like line
just beneath the physical singularity $r=0$ in the interior.  The horizon
also represents the singular curve in the re-scaled transformation
factor $\hat{W}_-$.  The incoming light-like surface represented by
the dashed blue line originating from the lower right region of the
diagram is the singular curve in the re-scaled transformation
factor $\hat{W}_+$.  The evolving radial mass scale $R_M(ct_a)$
is represented by the dashed blue curve originating at the
bottom corner of the diagram.  Each of these
dashed curves terminates at $(ct_a=0, r=0)$ when the
evaporation ceases, and cross indefinitely larger radial
coordinates in the distant past. 
The radial mass scale $R_M$ associated with the $(ct_a,r)$
coordinate anomaly in the highly curved metric
of the black hole geometry is clearly seen to differ from the
horizon coordinate $R_H$.  Indeed, $R_M (ct_a)$ cannot be a light-like
surface, since outgoing photons at this coordinate will momentarily be stationary,
while $R_M$ itself is steadily decreasing.  The behaviors of the
coordinate lines are clearly anomalous in the vicinity of the
radial mass scale $r \leq R_M$.  However, there is no physical singularity
introduced at this coordinate anomaly.

From the diagram, one can immediately determine that no photon emitted
from the region interior to the horizon $0< r <R_H$ can escape hitting
the physical singularity $r=0$. 
Similarly, as discussed in section \ref{horizonsection}, an
outgoing photon emitted from the region $R_H < r \leq  R_M$ is 
seen to escape hitting the singularity.  No incoming photon emitted
from the region to the right of the singular curve of the
re-scaled transformation factor $\hat{W}_+$ 
(the incoming dashed blue line)
\emph{can}
hit the physical singularity or communicate with the
interior region.
The horizon is always seen to lie within the radial mass scale $R_H < R_M$
as expected for the evaporating black hole. 
The physical singularity $r=0$, the horizon $R_H$ (which
is proportional to the radial mass scale), and the radial mass scale
$R_M$ are seen to vanish together, leaving a 
time-like curve $r=0$ associated with the final no
curvature Minkowski space-time, represented as the
upper triangular region in the diagram subsequent
to complete evaporation of the singularity.

\setcounter{equation}{0}
\section{Conclusions}
\indent

The large-scale geometry of a black hole of steadily decreasing mass scale
has demonstrated interesting aspects.  One might expect that a black
hole that has a horizon of infinite area as the time used by a distant observer
tends to $-\infty $ would
eventually contain any given observer within that horizon
at an early enough time.  The diagram in Fig. \ref{Penrose} clearly demonstrates
that a substantial region of space-time lies \emph{outside} of the regions of
significant curvature, even outside the region of causal in-going exchange
with the interior region of the black hole.  This region can be parameterized
in terms of multiples of the shrinking mass scale, always remaining external
to the interior regions.  The diagram also directly demonstrates the distinction
between the horizon and the radial mass scale for this dynamic geometry.

The diagram demonstrates some intriguing aspects of ``Asymptopia". 
As shown in Eq. \ref{Aofzeta}, the zero
at $\zeta=0$ in the temporal scale factor
$A(\zeta)$ that transforms the original coordinate $ct_a$
to the diagonal ``Schwarzschild-like" time
$ct_D$ at $r \rightarrow \infty $ dis-allows the use of the asymptotic
Schwarzschild geometry for correspondence of the 
Schwarzschild temporal parameter
with the diagonal time $ct_D$. 
This means that some aspects of the
asymptotic space-time described using the
coordinates $(ct_D,r)$ that corresponds to Schwarzschild space-time
at fixed $\zeta_o$ are not shared by corresponding asymptotic aspects of
that particular Schwarzschild space-time.  This
additionally manifests by the observation
that for static geometries, the form of $ct_D$ is ill-defined\cite{rivermodel}
in the asymptotic regime.

The ``radiations" associated with the evaporation of this black hole
are of a spatially coherent nature, vanishing on the space-like volume
$ct_a =0$.  Similar coherence is naturally found in the quantum nature
of interactions.  The dynamics for scalar quantum fields in this geometry
is being examined, and will be presented in a future submission. 
In addition, the authors are examining diagrams describing systems
that begin an accretion process and end a subsequent evaporation
process using modifications of the results presented here.  One
should be able to gain insight into the life cycle of a black hole
from these explorations.

\begin{center}
\textbf{Acknowledgments}
\end{center}
BAB would like to acknowledge the support of the NASA Administrator's
Fellowship Program.

\end{document}